\documentclass[seceq]{ptptex}

\usepackage{graphicx}
\title{%        %You can use \\ for explicit line-break
Can symmetric texture reproduce unitarity triangle and $m_b/m_{\tau}$ ? 
}

\author{%       %Use \scshape  for the family name
Masako \textsc{Bando $^{a,}$} \footnote{E-mail address: bando@aichi-u.ac.jp},
Satoru \textsc{Kaneko $^{b,}$} \footnote{E-mail address: satoru@phys.ocha.ac.jp},
Midori \textsc{Obara $^{c,}$} \footnote{E-mail address: bando@aichi-u.ac.jp}\\ and\\ 
Morimitsu \textsc{Tanimoto $^{d,}$} \footnote{E-mail address: tanimoto@muse.sc.niigata-u.ac.jp}
}

\inst{%     %Affiliation, neglected when [addenda] or [errata]
 $^a$ {\it Aichi University, Aichi 470-0296, Japan} \\[1mm] 
   $^b$  {\it Department of Physics, 
      Ochanomizu University, Tokyo 112-8610, Japan}  \\[1mm] 
   $^c$ {\it Institute of High Energy Physics,  Chinese Academy of Sciences, \\ 
   P.O. Box 918, Beijing 100049, China} \\[1mm] 
   $^d$  {\it Department of Physics, 
      Niigata University, Niigata,  950-2128, Japan}
}

\abst{%         %this abstract is neglected when [addenda] or [errata]
We study the SUSY SO(10) GUT with the symmetric two-zero textures of quark/lepton mass matrix 
which realize the Georgei-Jarlskog relations in down-type quark and charged lepton masses. 
We show that the important constraints to such framework come from the bottom-tau unification 
and the observed value of $\sin2\beta$, one of the angles in the CKM unitarity triangle. 
We investigate the symmetric two-zero textures assumed at the GUT scale by solving the MSSM 
renormalization group equations with  right-handed neutrino threshold effects. As a result, 
the value of $\tan\tilde{\beta}$ is constrained to be large enough and the representation of 
Higgs field which couples to the third generation of left- and right-handed neutrinos is 
restricted by the bottom-tau unification condition.   
The textures with vanishing 1-2 (and 2-1) element for up-type quarks and vanishing 1-3 (and 3-1) 
element for down-type quarks are favored by the observation of $\sin2\beta$. 
}

\begin{document}

\maketitle

%%%%%%%%%%%%%%%%%%%%%%% 
\section{Introduction} 
%%%%%%%%%%%%%%%%%%%%%%% 

One of the most successful predictions in supersymmetric grand unified theories (GUTs) is a gauge 
coupling unification at the GUT scale. 
On the other hand, 
any sufficient achievement of matter (Yukawa coupling) unification has not 
been completed yet and it is one of the most important tasks 
to explain the observed quark/lepton masses 
and mixings based on the GUT framework .   

Here we would like to notice that the well-known Georgi-Jarlskog (GJ) relations \cite{GJ} \begin{eqnarray} 
m_e=\frac{m_d}{3} 
\ ,\ \ \ 
m_\mu=3m_s 
\ ,\ \ \ 
m_\tau=m_b 
\ , 
\label{GJr} 
\end{eqnarray} 
can provide us 
with %made correction 
a guiding principle to construct a framework of matter unification. 
The above GJ relations at the GUT scale have  been known to 
explain 
the observed values of down quark and charged lepton masses at low energy quite successfully.   
Especially the first and second relations are quite interesting to 
reproduce the observed down and strange quark masses as well as $\mu $ and $e$ lepton masses 
in a unified way. Furthermore 
we notice that the ratio $\sqrt{m_d/(m_s+m_d)}$, which is $0.224\pm 0.004$ experimentally, 
is almost equal to the observed mixing value, $|V_{us}|_{\rm exp}=0.221-0.227$. 
This may be a strong indication of 
zero structure for 1-2 entry of the down-type quark mass matrix $M_d$, 
since the contribution of the mixing angle from the up-type quark mass matrix $M_u$, 
which shows more hierarchical structure, may be generally expected to be very small. 
We shall see this coincidence later. 
Thus a  natural setup for realizing the GJ relation is 
to introduce the zero texture of symmetric mass matrix with 
some vanishing entries. Such textures of quark mass matrix have been 
explored by many authors, for example, see refs.~\cite{RRR,watanabe}. 
The coefficient $-3$ in the GJ relations is originated from Clebsch-Gordan (CG) coefficients. 

On the other hand, the third relation in eq.~(\ref{GJr}), called bottom-tau  unification, 
should be addressed in more detail. Although the predicted value of $m_b/m_{\tau}$ 
at the low energy scale is qualitatively consistent with the experimental value, 
it depends quite strongly on the 3rd generation of right handed neutrino mass $M_{R3}$, as well as the strong coupling constant $\alpha_3$~\cite{btau}. 
Actually, it predicts a bit larger value than experimental values if $M_{R3}$ is well below the GUT scale. 
Furthermore, the recent results of the B factory, 
combined with the great advance in lattice QCD calculations, 
provide us with  much more precisely determined observable 
parameters in the CKM matrix. 
In particular the experimental value of $\sin2\beta$,
which is an angle in the CKM unitary triangle,
has been precisely determined in 
the last few years. This constrains the type of mass matrix textures very strictly. 
It is shown that the important constraints 
on fermion mass matrix textures come from the measurements of 
$\sin2\beta$~\cite{raby} and that the texture with zero 1-3 entry  cannot 
reproduce the experimental data. 

In the previous papers~\cite{BO,BKOT1}, we proposed a symmetric two-zero quark/lepton mass texture, 
which realizes the GJ relations and can successfully reproduce the observed bi-large mixings 
and mass-squared differences of neutrinos in the SUSY SO(10) GUT. 
In our framework, the predicted branching ratios of lepton flavor violation 
such as $\mu\to e\gamma$ are safely lower than the experimental upper bound and the 
observed baryon asymmetry of the universe can be explained through thermal 
leptogenesis scenario~\cite{BKOT2}. 
It is known (see for example, ref.\cite{btau}) that ratio of $m_b/m_{\tau}$ 
at the electroweak scale can be reproduced pretty well if $M_{R3}$ is of the GUT scale. 
This implies that the two zero texture of neutrino Dirac mass matrix $M_{{\nu}_D}$ in our model 
is preferable for predicting  $m_b/m_{\tau}$ because $M_{R3}$ is required 
to be almost of the GUT scale in our model, while the right-handed neutrino masses of the 1st and 2nd 
generations, $M_{R1}$ and $M_{R2}$, are very small. 
So, we expect that our model can reproduce not only  neutrino masses and mixing angles 
but also the down quark and charged lepton masses as well as quark mixing angles. 
However, our model adopted the texture with zero 1-3 
%%%%correction 
entry for both $M_u$ and $M_d$ 
%%%%%%% 
and it is not clear whether it would bring enough non-zero contribution 
to the 1-3 element 
%%correction 
to be 
%%%%%%%% 
consistent with the observed value of $\sin2\beta$. 
Therefore, 
the predicted values of $\sin2\beta$ and  $m_b/m_{\tau}$ 
%%% correction 
give a crucial information as to whether 
such matter unification scenarios with the GJ relations 
can be realized in nature, 
after 
%%%%%up to here 
taking account of the renormalization group equations (RGEs) running effect 
within the framework of unified bottom and tau Yukawa couplings. 
%%%%%%from her I made some correction%%%%%%%%%%%%%%%%%%%% 
Furthermore, 
%%%%%%%%%%%%%%%%%%%%%%%%%%%%% 
we think it is necessary to find  a nice framework 
of zero texture which is  consistent with the above current 
experimental data and at the same time which keeps the good 
GJ relations. 

The aim of this paper is to investigate the phenomenologically allowed symmetric two-zero 
quark/lepton mass matrix textures 
in the SUSY SO(10) GUT framework with the current experimental bounds of  $\sin2\beta$ and  $m_b/m_{\tau}$. 
We make numerical calculation of the observed values, especially $m_b/m_{\tau }$ and $\sin2\beta$, 
by solving the RGEs with right-handed neutrino threshold effects. 
We also show the constraints to $M_{R3}$ and $\tan\tilde \beta$. 
\footnote{We write here  $\tan\tilde \beta$ for the notation of the parameter 
expressing the ratio of VEVs of up and down Higgs fields to 
discriminate it from  $\sin2\beta$.} 

This paper is organized as follows. 
In the next section, we present the textures of quark/lepton mass matrix which we adopt in this paper. 
In the third section, we show the results of numerical calculations. 
The final section is devoted to the discussion of this paper.   

%%%%%%%%%%%%%%%%%%%%%%% 
\section{Symmetric two-zero textures} 
%%%%%%%%%%%%%%%%%%%%%%% 

We consider the SUSY SO(10) GUT with the {\bf 10} and {\bf 126} representation of Higgs multiplets. 
The SO(10) gauge group is assumed to be broken down to the standard model gauge group through the 
Pati-Salam symmetry. 
In the above setup,   
the up and down quark mass matrices : $M_u$, $M_d$,  charged lepton and neutrino Dirac mass matrices : 
$M_e$, $M_{\nu_D}$ are symmetric ones. 
Many authors have investigated the two-zero textures in the above setup~\cite{Chen,two-zero3}. 
We have assumed that either the {\bf 10} or {\bf 126} Higgs multiplet dominantly couples to the fermions 
in each generations.   
Consequently, 
we obtain the relation among mass matrices : $M_u=M_{\nu_D}$ and $M_d=M_e$ except for the CG 
coefficients $1$ or $-3$, which can appear in some elements in $M_e$ and $M_{\nu_D}$. 
The up quark and neutrino Dirac mass matrices are taken to be the following textures 
\footnote{ 
In addition to the texture U1 which we have adopted in our previous papers~\cite{BO,BKOT1,BKOT2}, 
we here also  introduce texture U2. The latter  texture is obtained by exchanging 
the 2nd and 3rd generation suffices, and it is expected not to make so much difference 
at least to the calculated neutrino masses and mixings because in the neutrino sector 
the 2nd and 3rd generations are almost maximally. Nevertheless, 
we guess that it does bring much difference to quark sector. 
} : 
\begin{eqnarray} 
&& 
{\rm U1}\ \     
M_u= 
\left( 
\begin{array}{ccc} 
 &   a_u&   \\ 
 a_u& b_u  & c_u  \\ 
 &  c_u & d_u   
\end{array} 
\right) 
\ \ {\rm and}\ \ 
M_{\nu_D}= 
\left( 
\begin{array}{ccc} 
 & x_{12}a_u &     \\ 
 x_{12}a_u & x_{22}b_u  & x_{23}c_u  \\ 
 &  x_{23}c_u & x_{33}d_u   
\end{array} 
\right) 
\ , 
\\ 
&& 
{\rm U2}\ \     
M_u= 
\left( 
\begin{array}{ccc} 
 &  & a_u   \\ 
 & b_u  & c_u  \\ 
 a_u&  c_u & d_u   
\end{array} 
\right) 
\ \ {\rm and}\ \ 
M_{\nu_D}= 
\left( 
\begin{array}{ccc} 
 & &x_{13}a_u      \\ 
  & x_{22}b_u  & x_{23}c_u  \\ 
 x_{13}a_u&  x_{23}c_u & x_{33}d_u   
\end{array} 
\right) 
\ ,
\end{eqnarray} 
where $a_{u}, b_{u}, c_{u}$ and $d_{u}$ are complex numbers, 
and CG coefficients $x_{ij}$ can be taken as $1$ or $-3$. 
We denote the above assumptions of textures as U1 and U2 hereafter. 
We have adopted the texture U1 in refs.~\cite{BO,BKOT1,BKOT2}. 
In this paper, we take $x_{12}=x_{13}=x_{22}=x_{23}=1$, 
because the values of $x_{ij}$ except for $x_{33}$ do not affect to the RGEs running of the observables 
in quark and charged lepton sector. 
The constraint on $x_{33}$ from the bottom-tau unification will be discussed in the next section 
by showing numerical results. 

The down quark and charged lepton mass matrices are given by 
\begin{eqnarray} 
&& 
{\rm D1}\ \     
M_d= 
\left( 
\begin{array}{ccc} 
 &   a_d&   \\ 
 a_d& b_d  & c_d  \\ 
 &  c_d & d_d   
\end{array} 
\right) 
\ \ {\rm and}\ \ 
M_e= 
\left( 
\begin{array}{ccc} 
 & a_d &     \\ 
 a_d & -3b_d  & c_d  \\ 
 &  c_d & d_d   
\end{array} 
\right) 
\ , 
\end{eqnarray} 
where $a_{d}, b_{d}, c_{d}$ and $d_{d}$ are also complex numbers. 
The CG coefficient in 2-2 element of $M_e$ is the crucial ingredient to realize the GJ relations. 
In addition to the above texture, 
we also consider the following three-zero texture for the down quark and the charged lepton mass matrices : 
\begin{eqnarray} 
&& 
{\rm D1'}\ \     
M_d= 
\left( 
\begin{array}{ccc} 
 &   a_d&   \\ 
 a_d& b_d  &   \\ 
 &   & d_d   
\end{array} 
\right) 
\ \ {\rm and}\ \ 
M_e= 
\left( 
\begin{array}{ccc} 
 & a_d &     \\ 
 a_d & -3b_d  &   \\ 
 &   & d_d   
\end{array} 
\right) 
\ ,   
\end{eqnarray} 
which is a special case of D1, imposing further zero to the 2-3 element 
\footnote{ 
It might be possible to introduce the case, in which we impose further zeros  to 
the matrix elements of $M_u$.  However, we have already recognized that 
it does not reproduce the neutrino bi-large mixing angles any more and 
the minimum parameter set should not be less than 4 for $M_{\nu_D}$. 
Thus the textures, U1 and U2 with D1 and D1', are natural extension of our previous model~\cite{BO,BKOT1,BKOT2}   
}. 
The texture D1$'$ were originally proposed by Georgei and Jarlskog~\cite{GJ} and also investigated in ref~\cite{RRR}. 

The right-handed neutrino mass matrix is taken to be 
\begin{eqnarray} 
M_R= 
\left( 
\begin{array}{ccc} 
 &   a_R&   \\ 
 a_R&   &   \\ 
 &   & d_R   
\end{array} 
\right)\ , 
\end{eqnarray} 
where we assume $|a_R|\ll |d_R|=M_{R3}\simeq M_{\rm GUT}$ where $M_{R3}$ is the third generation of 
right-handed neutrino mass. 
There are two characteristic features of this form. 
First, it reproduces the observed bi-large mixings of neutrinos~\cite{BO,BKOT1}. 
Second, it predicts almost equal Majorana masses for the 1st and 2nd generations of right-handed neutrinos. 
The former is important to derive $m_\tau/m_b$ ratio, 
while in the latter, 
degenerate masses guarantee the baryon asymmetry of the universe through leptogenesis~\cite{FuYa} in taking account of the resonance effects~\cite{pilaftsis}.   
The left-handed Majorana neutrino mass matrix is obtained by the seesaw mechanism : 
$M_\nu=-M_{\nu_D}^T M_R^{-1} M_{\nu_D}$. 

%%%%%%%%%%%%%%%%%%%%%%% 
\section{Numerical analysis} 
%%%%%%%%%%%%%%%%%%%%%%% 

In this section, 
we show the predicted values of the observable parameters in the CKM matrix : $V_{\rm CKM}$, 
the quark and charged lepton masses at the electroweak scale by solving RGEs with the right-handed 
neutrino threshold effects~\cite{RGEs}. 

\begin{figure}[t]
\centerline{
\includegraphics[width=8.2 cm,height=6.5 cm]{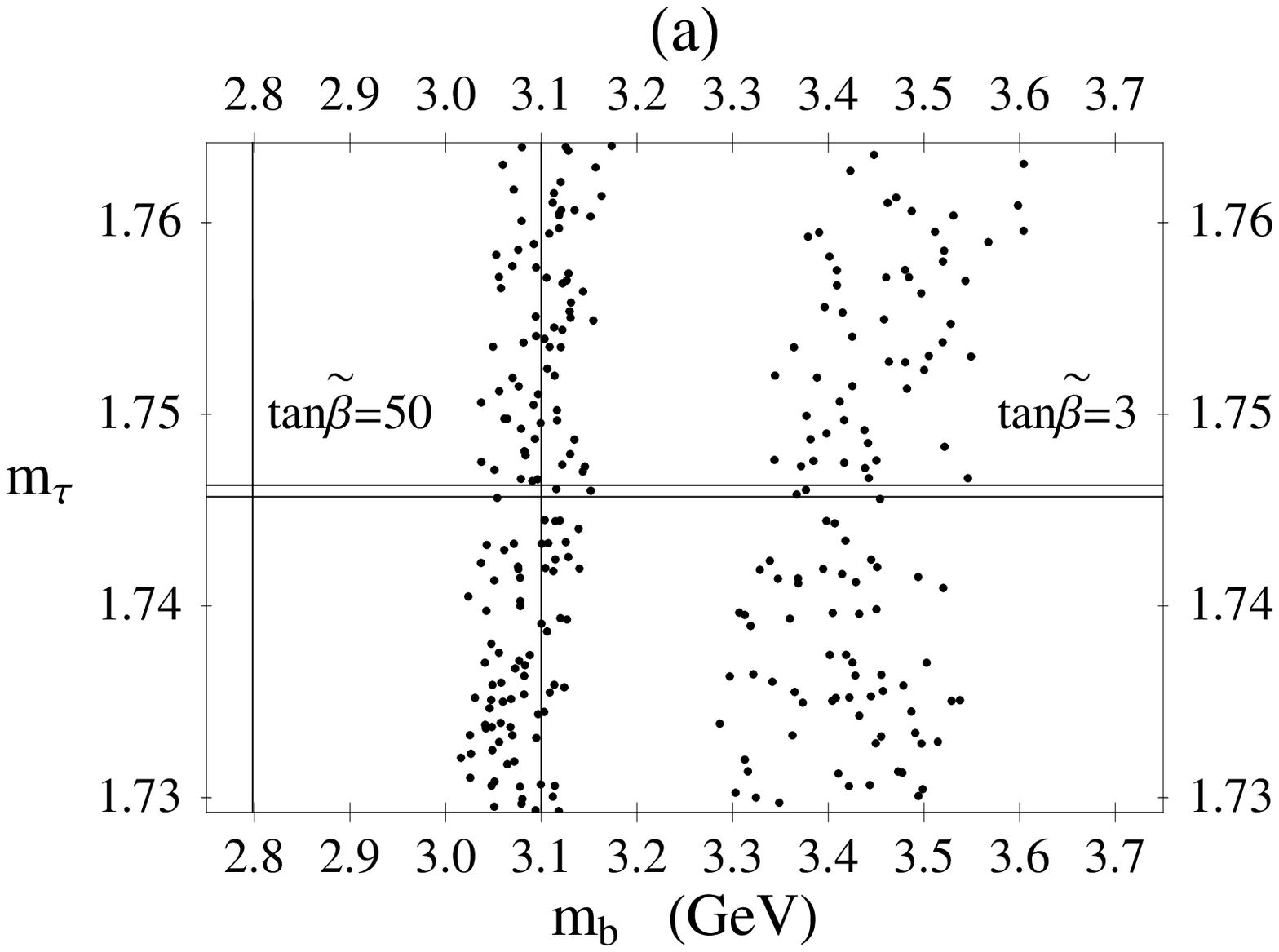}
\includegraphics[width=8.2 cm,height=6.5 cm]{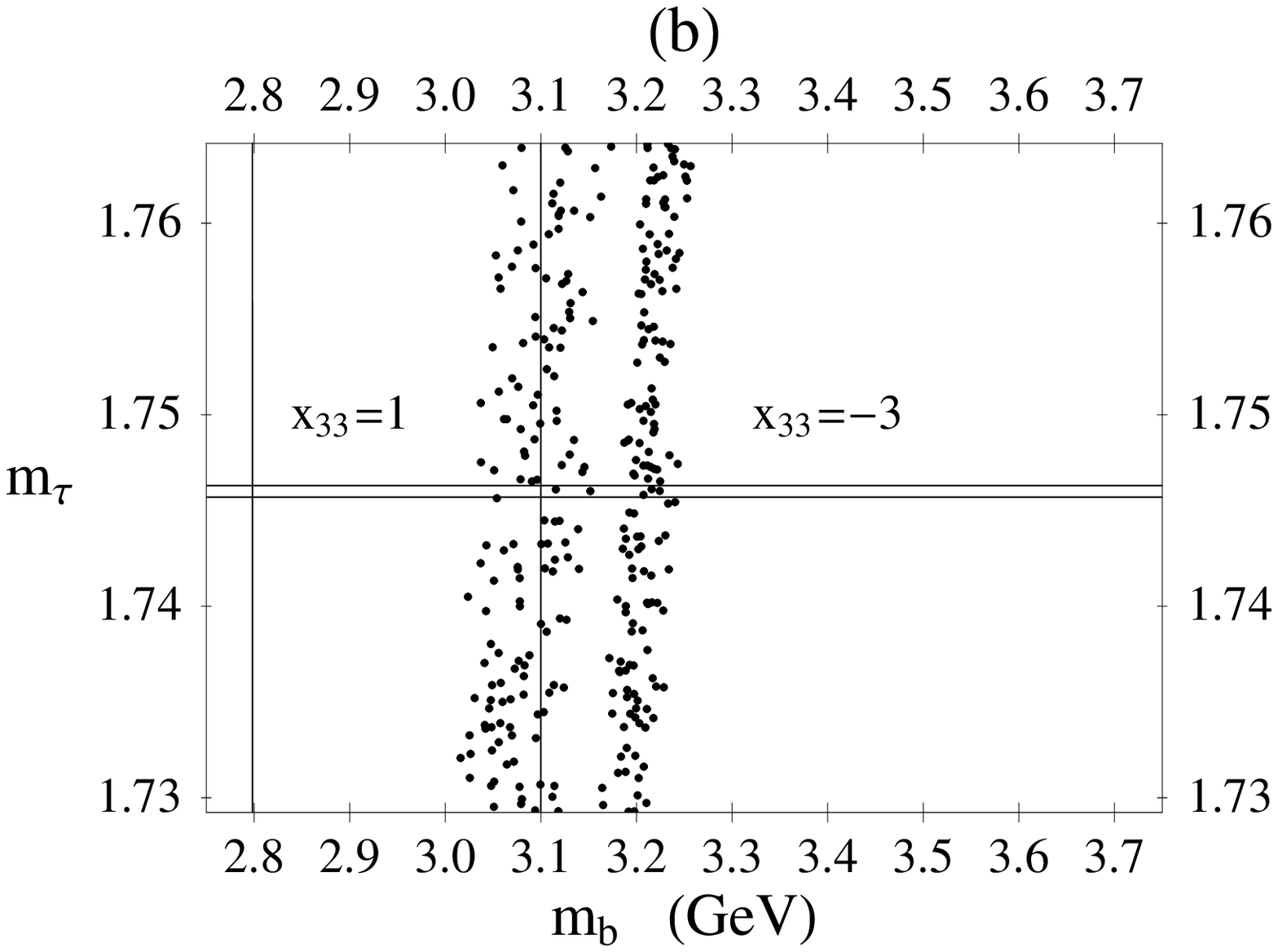}
}
\caption{
(a): The scattered plots of $m_b$ and $m_\tau$ in the case of U1-D1 textures for $\tan\tilde{\beta}=3$ and $50$. 
The $x_{33}$ is fixed as $x_{33}=1$ in this case. 
(b): The case of $x_{33}=1,-3$ are shown when $\tan\tilde{\beta}$ is fixed as $\tan\tilde{\beta}=50$. 
In  both figures, 
the right-handed neutrino masses are fixed as $M_{R1}=M_{R2}=10^{10}$ GeV and $M_{R3}=10^{15}$ GeV. 
The two horizontal lines in the figures correspond to the allowed range of $m_\tau$.
The vertical lines correspond to the allowed range of $m_b$. 
We find that the case of large $\tan\tilde{\beta}$, say $\tan\tilde{\beta}\simeq 50$ and $x_{33}=1$ is favored by bottom-tau unification comparing with the experimentally allowed range for bottom quark mass : $2.8<m_b<3.1$ GeV from these figures. 
}
\label{btauU1D1} 
\end{figure}

The procedure that we adopt here for generating scattered plots of Fig.~\ref{btauU1D1}, 
Fig.~\ref{U1D1}, Fig.~\ref{U2D1} and Fig.~\ref{U2D2} is explained as follows.   
First, we generate random numbers for $a_{u,d},b_{u,d},c_{u,d},d_{u,d}$ in the mass matrices. 
Then, after solving the RGEs for Yukawa couplings with the generated values of $a_{u,d},b_{u,d},c_{u,d},d_{u,d}$ 
at the GUT scale, 
we plot the observable parameters if the up quark and charged lepton masses fall into the following 
experimentally allowed range : 
\begin{eqnarray} 
m_u(M_Z)=0.8 - 3.0\ {\rm MeV},& 
m_c(M_Z)=500 - 800\ {\rm MeV},&   
m_t(M_Z)=170 - 180\ {\rm GeV}, 
\nonumber\\ 
m_e(M_Z)=0.487\ {\rm MeV},&   
m_\mu(M_Z)= 103\ {\rm MeV},&   
m_\tau(M_Z)= 1.75\ {\rm GeV},   
\nonumber\\ 
|V_{us}|=0.221 - 0.227\ ,&&   
\label{ex1} 
\end{eqnarray} 
where the $m_u$, $m_c$ and $m_t$ are taken to have maximally wide range \cite{PDG}, 
the charged lepton masses are allowed to have 1\% error around the displayed central value~\cite{KF}. 
We then obtain the predicted values of $m_d$, $m_s$, $m_b$, $|V_{ub}|$, $|V_{cb}|$ and $\sin2\beta$. 
The another important observable is the strong coupling constant $\alpha_3$ which is defined 
as $\alpha_3\equiv g_3^2/4\pi$ : 
\begin{eqnarray} 
\alpha_3(M_Z)=0.1187\pm 0.0020\ , 
\label{a3} 
\end{eqnarray} 
where we use the value in~\cite{PDG}. 
As we will see later soon, 
the low-energy value of the bottom to tau lepton mass ratio depends on the value of $\alpha_{3}$.

\begin{figure}[t]
\centerline{
\includegraphics[width=7.2 cm,height=7.2 cm]{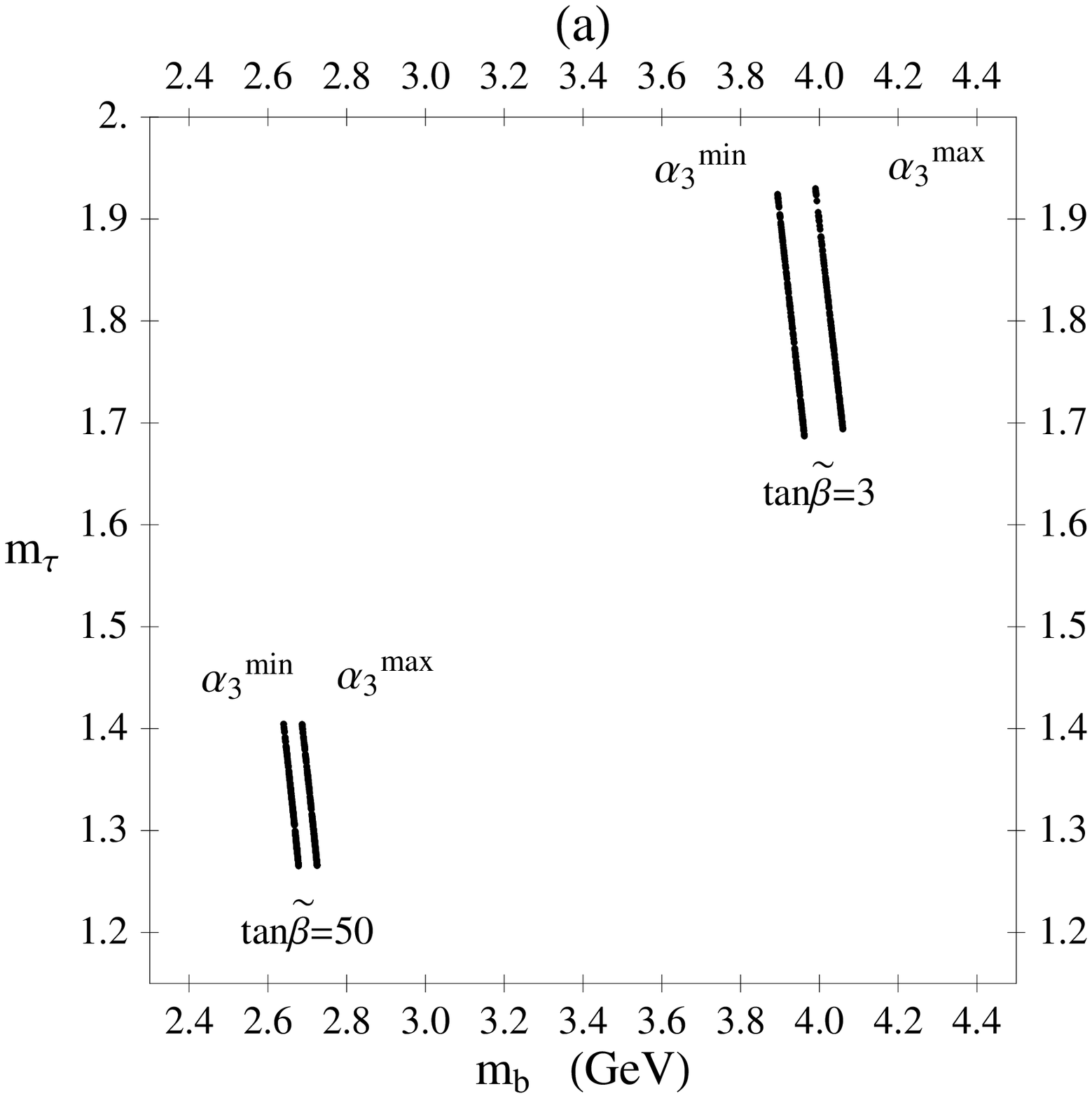}
\includegraphics[width=7.2 cm,height=7.2 cm]{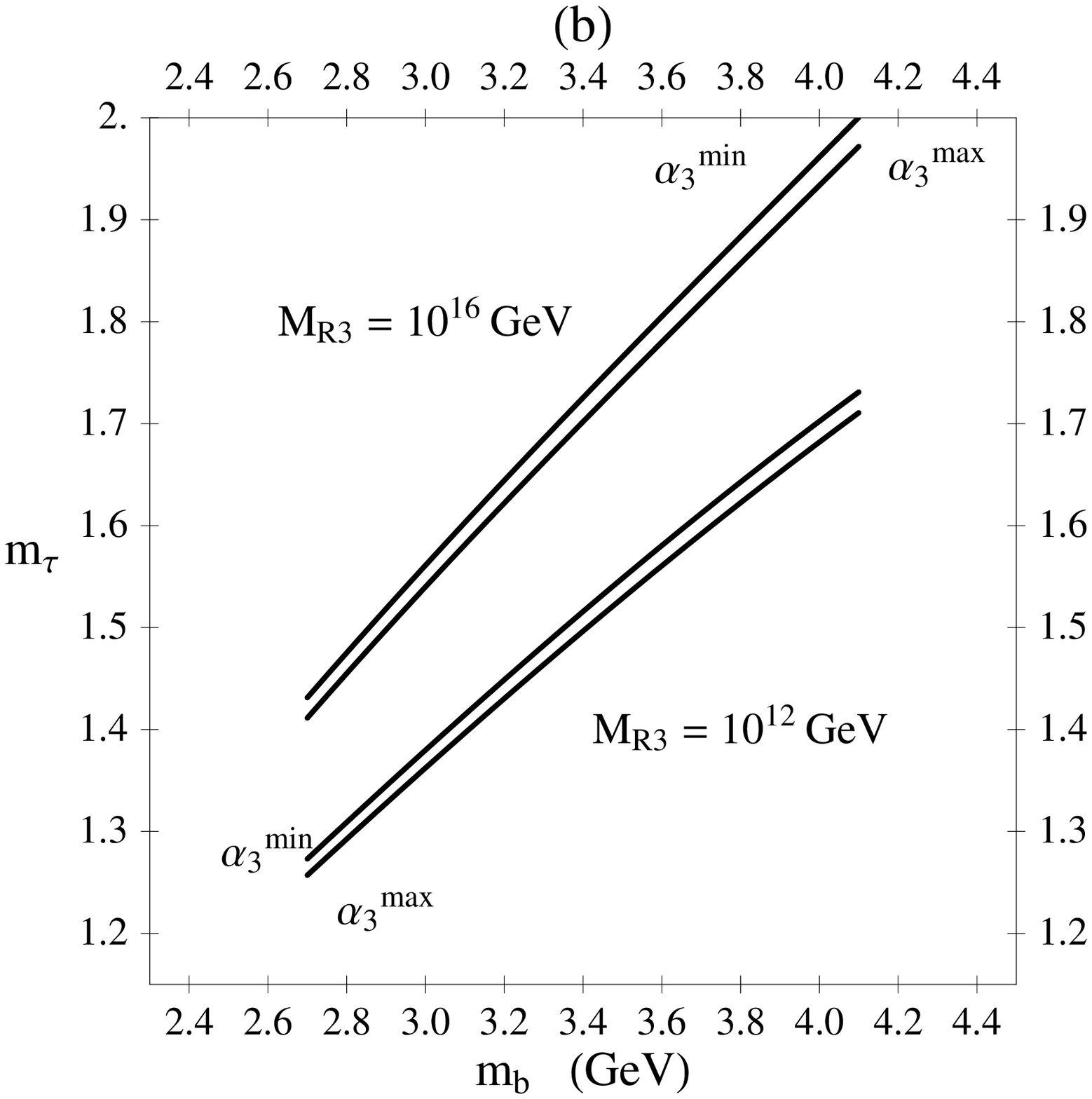}
}
\caption{
(a): The $\alpha_3$ and $\tan\tilde{\beta}$ dependence of $m_b$ and $m_{\tau}$ are shown. 
The heaviest right-handed neutrino mass is taken to have the range of  $M_{R3}=10^{12}\sim 10^{16}$ GeV. 
(b): The $\alpha_3$ and $M_{R3}$ dependence of $m_b$ and $m_{\tau}$ are shown. 
The $\tan\tilde{\beta}$ is taken to have the range of  $\tan\tilde{\beta}=3\sim 50$. 
The minimal and maximal value of $\alpha_3$ are taken according to eq.(\ref{a3}) in both  (a) and (b), respectively. 
The lighter right-handed neutrino masses are fixed as $M_{R1}=M_{R2}=10^{10}$ GeV in these figures. 
}
\label{btautbmr} 
\end{figure}

%%%%%%%%%%%%%%%%%%%%%%% 
\subsection{bottom-tau unification} 
%%%%%%%%%%%%%%%%%%%%%%% 
Here, we discuss the constraint which comes from bottom-tau unification. 
Before showing the numerical results, 
we present the parameter dependence of the bottom to tau mass ratio at the low-energy scale. 
In the MSSM with heavy right-handed neutrinos, 
the 1-loop RGE for the ratio of bottom and tau masses is approximately given by 
\begin{eqnarray} 
\frac{d}{dt} 
\left(\frac{m_b}{m_\tau}\right) 
\simeq 
\frac{1}{16\pi^2} 
\left(\frac{m_b}{m_\tau}\right) 
\left[ 
(y_u)_{33}^2-(y_\nu)_{33}^2 
+ 
3\left( 
(y_d)_{33}^2-(y_e)_{33}^2 
\right) 
- 
\left( 
\frac{16}{3}g_3^2-\frac{4}{3}g_1^2 
\right) 
\right]\ ,\nonumber\\ 
\label{RGE} 
\end{eqnarray} 
where $g_{1,3}$ are the gauge couplings of the standard model gauge group, 
$y_{u,d,\nu,e}$ are the Yukawa couplings for the up and down quarks, neutrinos and charged leptons, 
respectively, 
and $t\equiv \ln\mu$. 
The top quark Yukawa coupling $(y_u)_{33}$ tends to cancel the 
contribution of the  $g_3$ term on the right hand side in eq.(\ref{RGE}). 
This cancellation is important to reproduce the phenomenologically allowed low-energy value of 
the bottom to tau mass ratio. 
Since the tau neutrino Yukawa coupling $(y_\nu)_{33}$ cancels the top quark one, 
the bottom quark mass becomes smaller if the absolute value of the 
CG coefficient $x_{33}$ in 3-3 element decreases. 
Moreover, the lower values of $\tan\tilde{\beta}$ and $M_{R3}$ which 
determines a decoupling scale of the third generation of right-handed neutrino 
are also disfavored for the same reason. 
The explicit example of plots in $m_b-m_\tau$ plane are shown in Fig.~\ref{btautbmr} 
for the different values of $\alpha_3$, $M_{R3}$ and $\tan\tilde{\beta}$, respectively. 
In these figures, the RGEs are numerically solved.
The $\alpha_3$ and $\tan\tilde{\beta}$ dependence of $m_b$ and $m_{\tau}$ are shown in Fig.~\ref{btautbmr}(a). 
The heaviest right-handed neutrino mass is taken to have the range of  $M_{R3}=10^{12}\sim 10^{16}$ GeV in Fig.~\ref{btautbmr}(a). 
The $\alpha_3$ and $M_{R3}$ dependence of $m_b$ and $m_{\tau}$ are shown in Fig.~\ref{btautbmr}(b). 
The $\tan\tilde{\beta}$ is taken to have the range of  $\tan\tilde{\beta}=3\sim 50$. 
The minimal and maximal values of $\alpha_3$ are taken according to eq.~(\ref{a3}) 
in Fig.~\ref{btautbmr}(a) and \ref{btautbmr}(b), respectively. 
It is shown that both $m_b$ and $m_\tau$ are proportional to $\tan\tilde{\beta}$. 
Furthermore, $m_b$ ($m_\tau$) is also proportional to $\alpha_3$ ($M_{R3}$), respectively. 
We can confirm the above arguments 
by the figures of the results of  numerical calculations. 

Fig.~\ref{btauU1D1} shows the scattered plots in $m_b-m_\tau$ plane in the case of U1-D1 
for the different values of $\tan\tilde{\beta}$ and $x_{33}$ in taking account of the constraints shown in eqs.~(\ref{ex1}) and (\ref{a3}) except for $m_\tau$. 
In this analysis, the right-handed neutrino masses are fixed as $M_{R1}=M_{R2}=10^{10}$ GeV 
and $M_{R3}=10^{15}$ GeV, respectively. 
In the Fig.~\ref{btauU1D1}(a), we take $x_{33}=1$. 
We find that large $\tan\tilde{\beta}$ is favored by the bottom-tau unification with allowed range 
of bottom quark mass : $2.8<m_b<3.1$ GeV. 
The Fig. \ref{btauU1D1}(b) shows the $x_{33}$ dependence  of bottom quark mass. 
We easily find that the allowed values of bottom quark mass are obtained in the case of $x_{33}=1$. 
Therefore, we will consider only the case of 
\begin{eqnarray} 
x_{33}=1\ \ \ {\rm and}\ \ \ \tan\tilde{\beta}=50\ , 
\end{eqnarray} 
hereafter in this paper. 

Our next task is to show the difference of the down and strange quark mass predictions in the textures of mass matrices. 
Since the low energy value of the bottom to tau mass ratio is not sensitive to the textures of 
the up quark and neutrino Dirac mass matrices, 
it is enough to show the difference in the case of D1 and D1$'$. 
For the case of U1-D1 textures with $x_{33}=1$ and $\tan\tilde{\beta}=50$, 
the down and strange quark masses are predicted to be 
\begin{eqnarray} 
m_d(M_Z)=3.7-5.0\ {\rm MeV}\ ,\ \ m_s(M_Z)=55-74\ {\rm MeV} 
\ ,
\end{eqnarray} 
where the constraints of eqs.(\ref{ex1}) and (\ref{a3}) are considered, respectively.
Almost the same values are predicted in the case of U2-D1. 
These values are nicely located within the experimental allowed region : 
$m_d(M_Z)= 2.6-5.2$ MeV and $m_s(M_Z)= 52-85$ MeV \cite{watanabe,KF}. 
For the case of U1-D1$'$ textures with $x_{33}=1$ and $\tan\tilde{\beta}=50$, 
the down and strange quark masses are predicted to be 
\begin{eqnarray} 
m_d(M_Z)=3.2-3.4\ {\rm MeV}\ ,\ \ m_s(M_Z)=81-85\ {\rm MeV} 
\ . 
\end{eqnarray} 
Almost the same values are predicted in the case of U2-D1$'$. 
These values are more strongly restricted rather than the case of D1. 
We recognize that these two textures of down quark mass matrix, $M_d$, 
can reproduce the observed down and strange quark masses. 

%%%%%%%%%%%%%%%%%%%%%%% 
\subsection{The CKM matrix and unitarity triangle} 
%%%%%%%%%%%%%%%%%%%%%%% 

\begin{figure}[t]
\centerline{
\includegraphics[width=7.2 cm,height=7.2 cm]{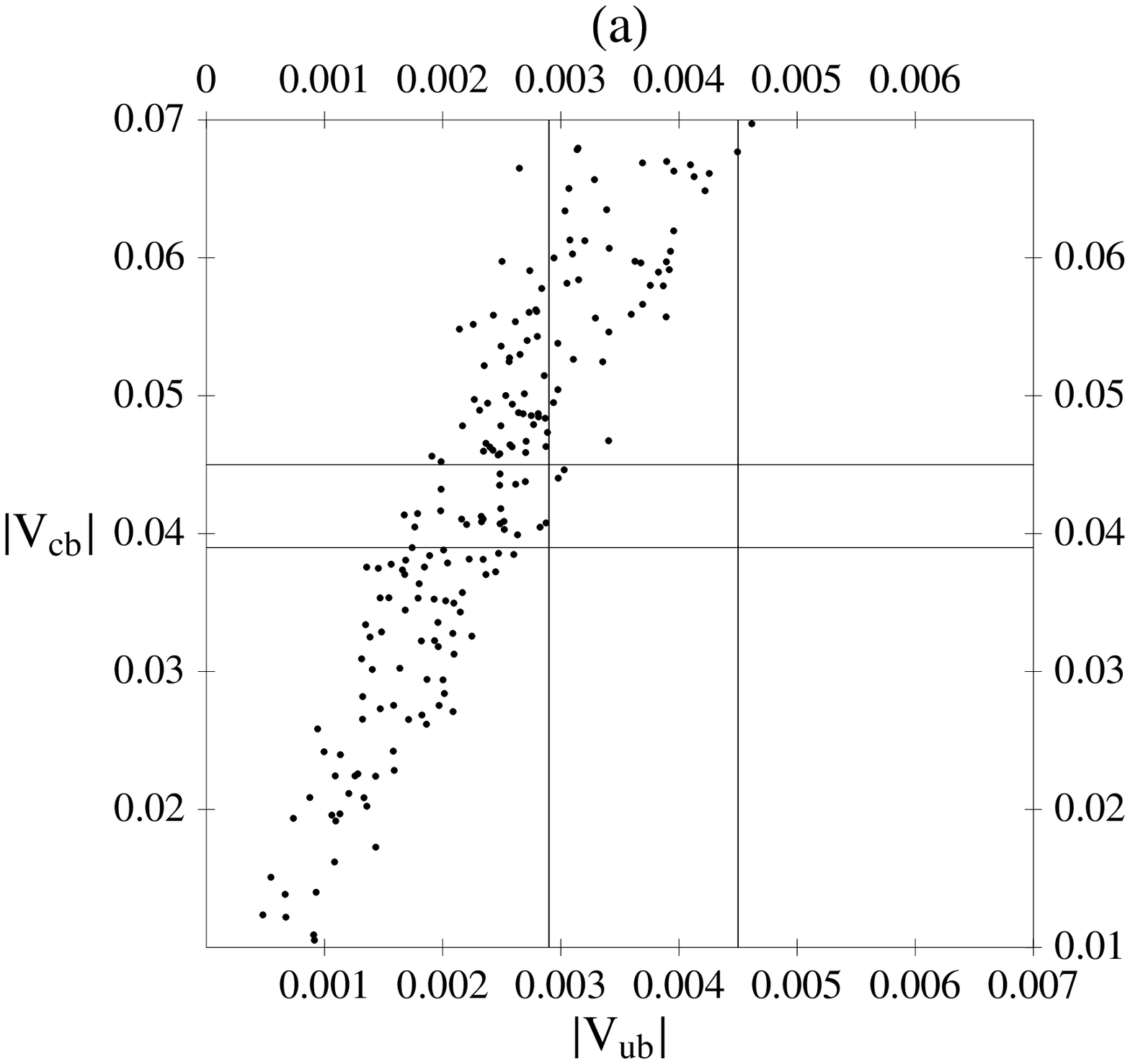}
\includegraphics[width=7.2 cm,height=7.2 cm]{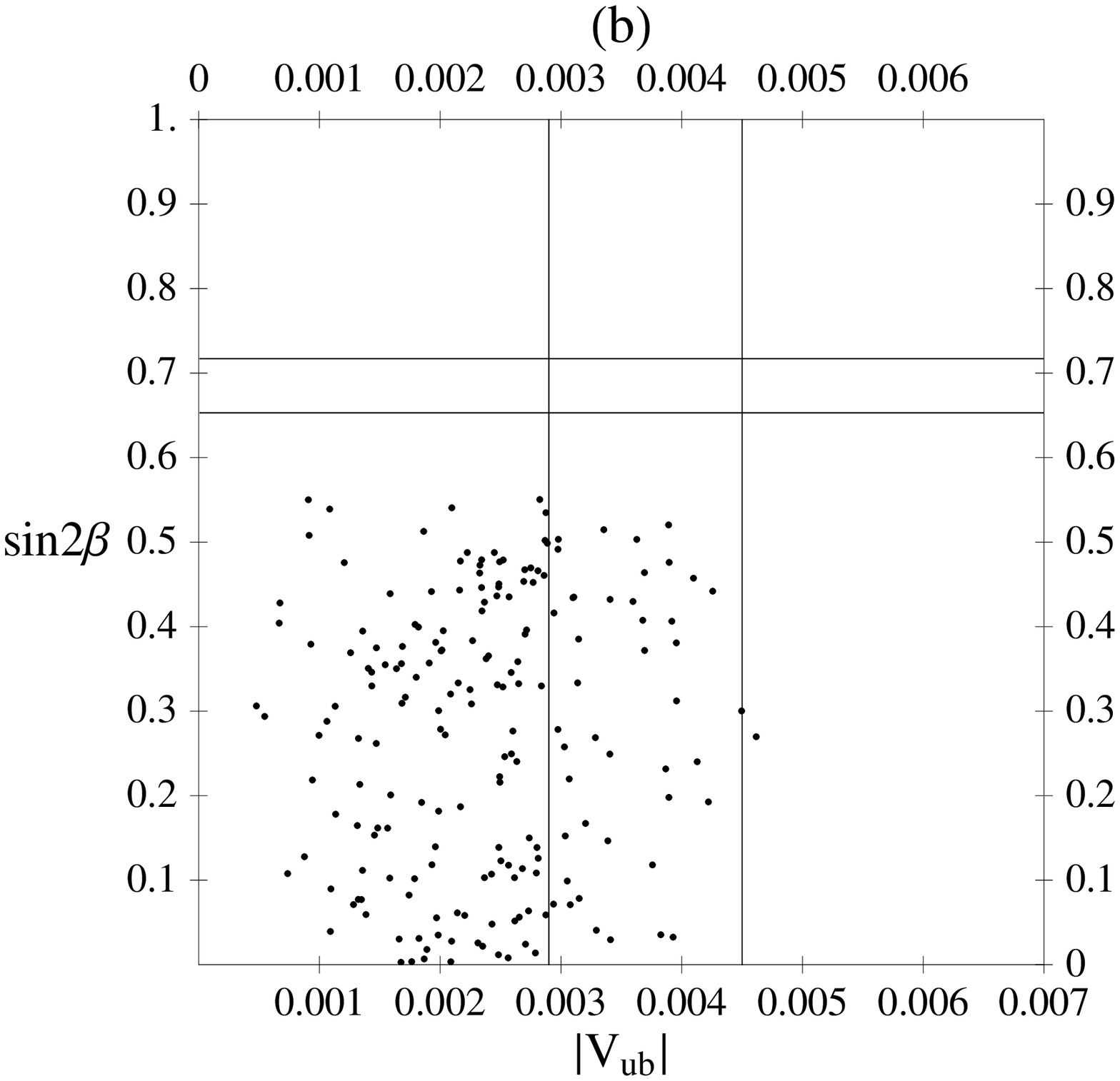}
}
\caption{
(a): Predicted values of $|V_{cb}|$ are plotted as a function of $|V_{ub}|$ for the case of U1-D1. 
(b): Predicted values of $\sin2\beta$ are plotted as a function of $|V_{ub}|$ for the case of U1-D1.
In both figures, the constraints shown in eqs.(\ref{ex1}) and (\ref{a3}) are considered. 
The relevant parameters are fixed as $\tan\tilde{\beta}=50$, $x_{33}=1$, $M_{R1}=M_{R2}=10^{10}$ GeV 
and $M_{R3}=10^{15}$ GeV, respectively. 
The two horizontal lines in the figures correspond to the allowed range of $|V_{cb}|$ and $\sin2\beta$, respectively. 
The vertical lines correspond to the allowed range of $|V_{ub}|$. 
}
\label{U1D1} 
\end{figure}

\begin{figure}[t]
\centerline{
\includegraphics[width=7.2 cm,height=7.2 cm]{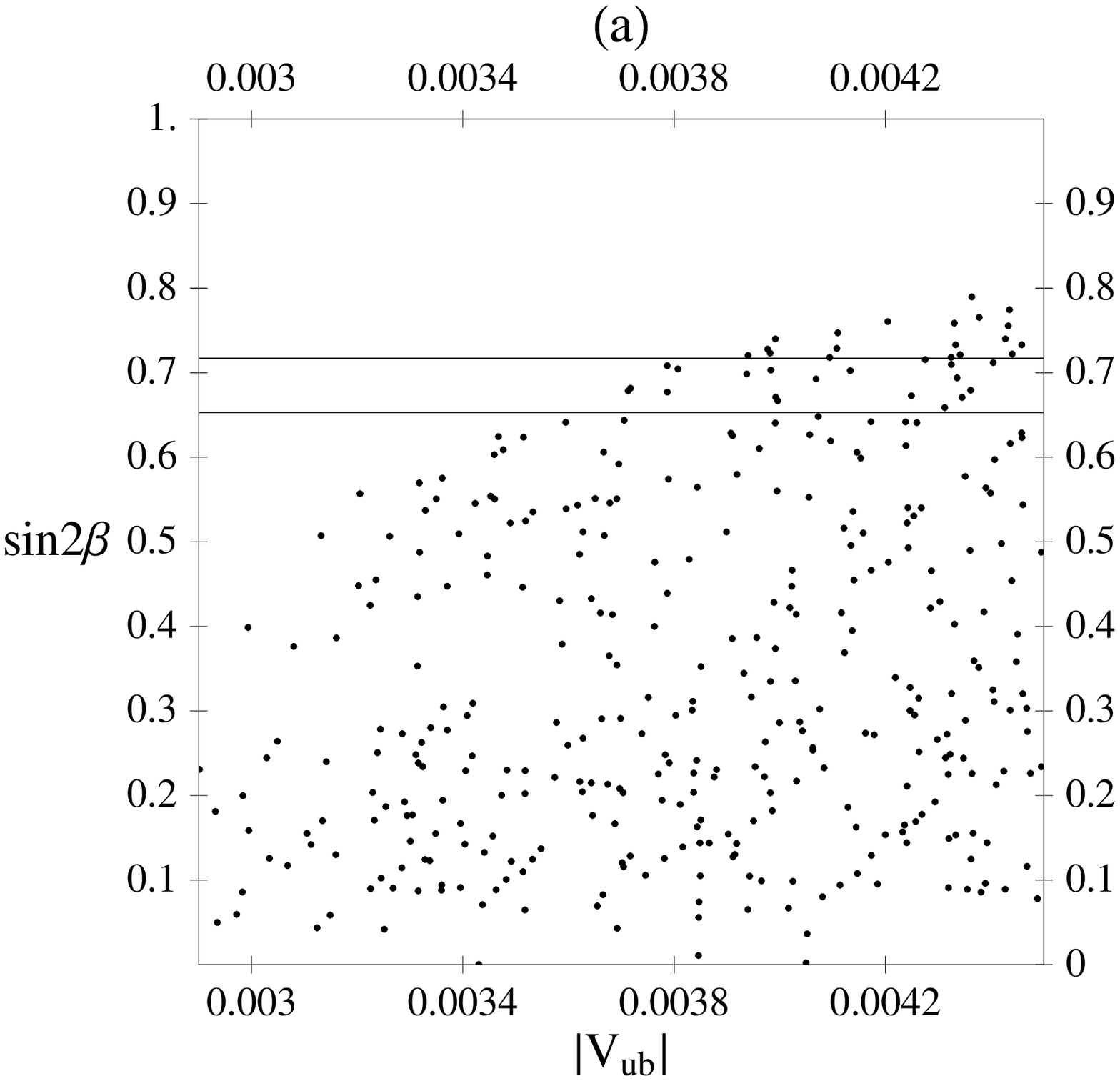}
\includegraphics[width=7.2 cm,height=7.2 cm]{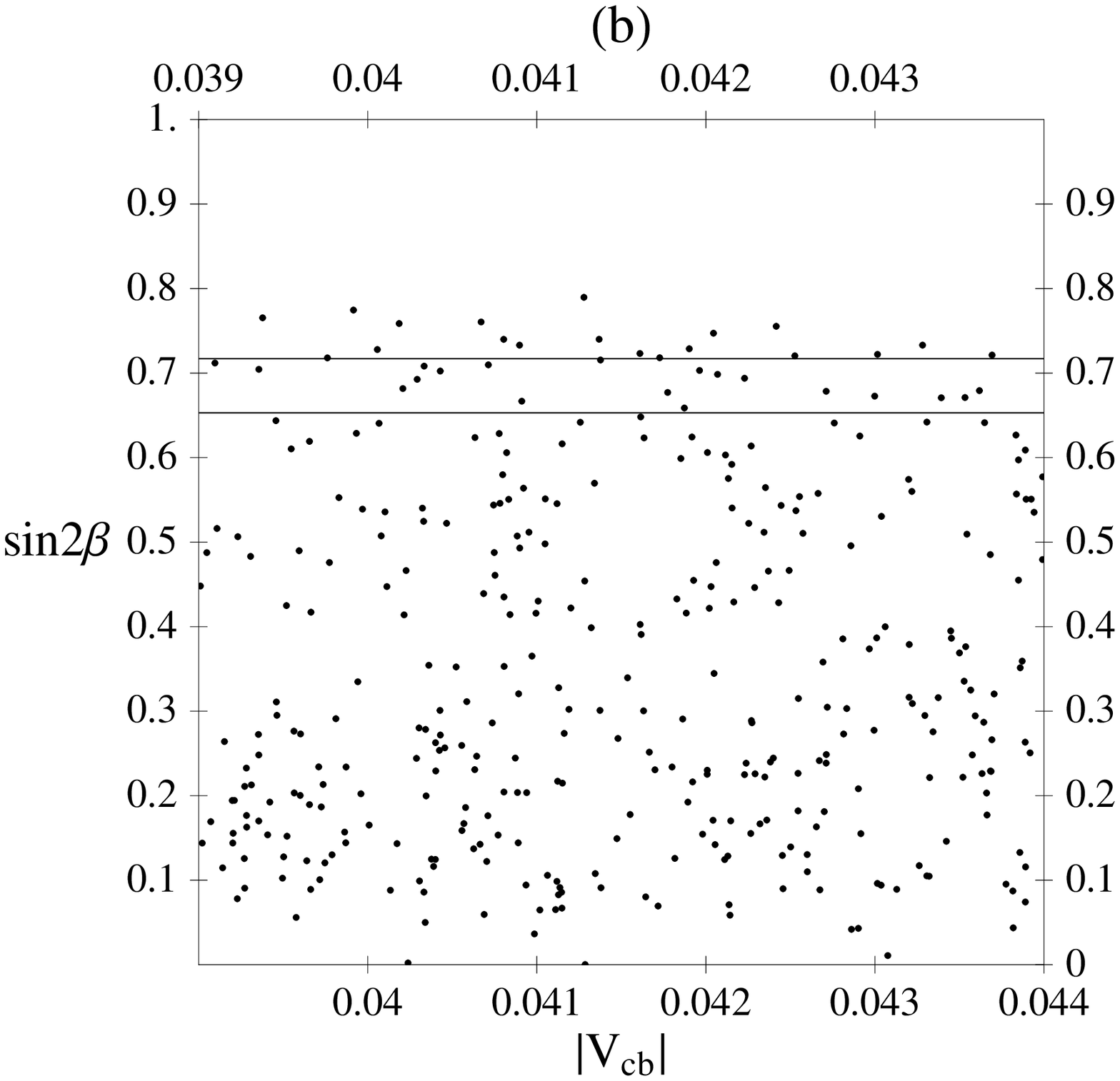}
}
\caption{(a): Predicted values of $\sin2\beta$ are plotted as a function of $|V_{ub}|$ for the case of U2-D1. 
(b): Predicted values of $\sin2\beta$ are plotted as a function of $|V_{cb}|$ for the case of U2-D1. 
In both figures, the constraints shown in eqs.(\ref{ex1}), (\ref{a3}) and (\ref{ex2}) are considered. 
The relevant parameters are fixed as $\tan\tilde{\beta}=50$, $x_{33}=1$, $M_{R1}=M_{R2}=10^{10}$ GeV and $M_{R3}=10^{15}$ GeV, respectively. 
The two horizontal lines in the figures correspond to the allowed range of $\sin2\beta$. 
}
\label{U2D1} 
\end{figure}

\begin{figure}[t]
\centerline{
\includegraphics[width=7.2 cm,height=7.2 cm]{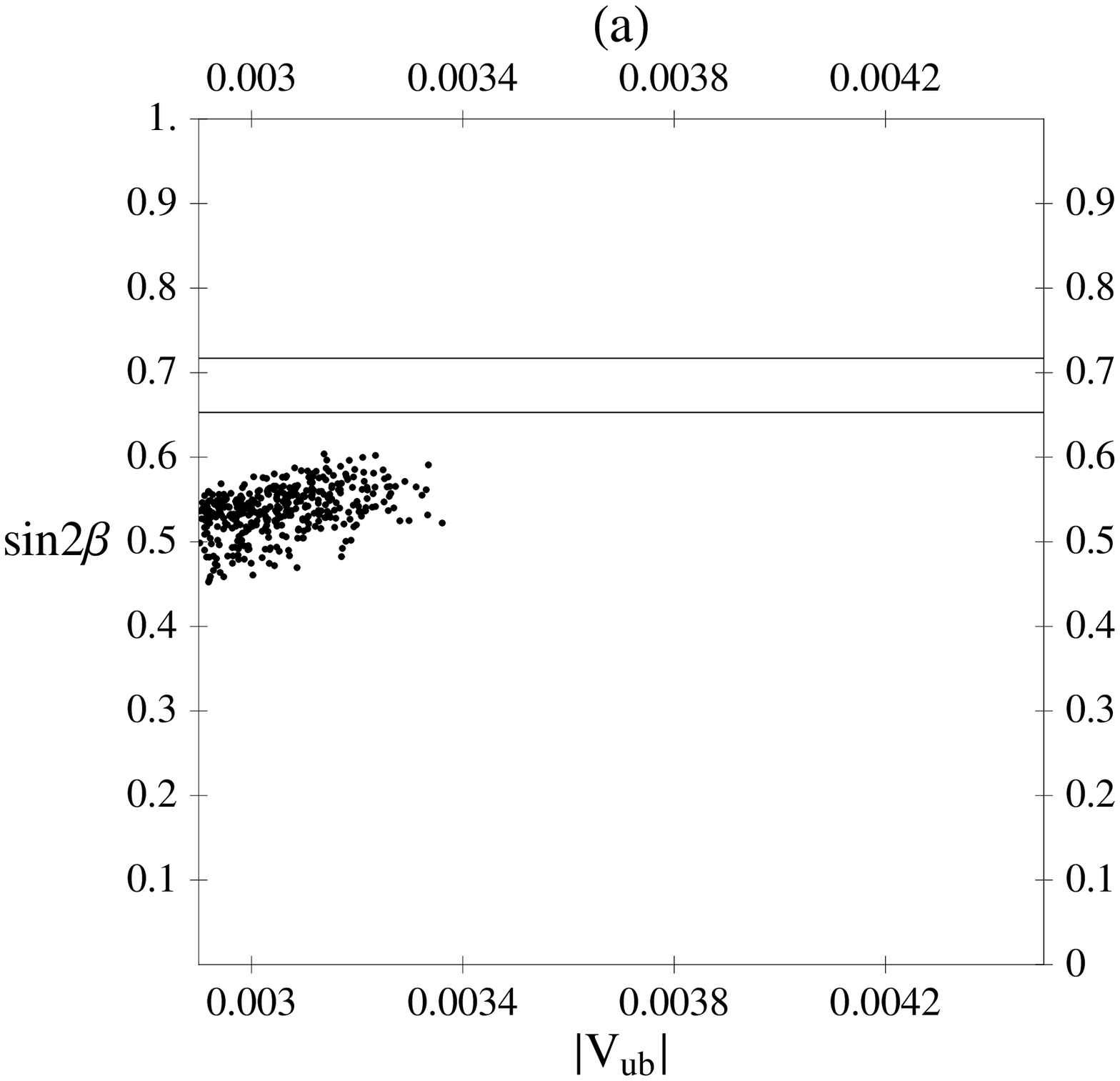}
\includegraphics[width=7.2 cm,height=7.2 cm]{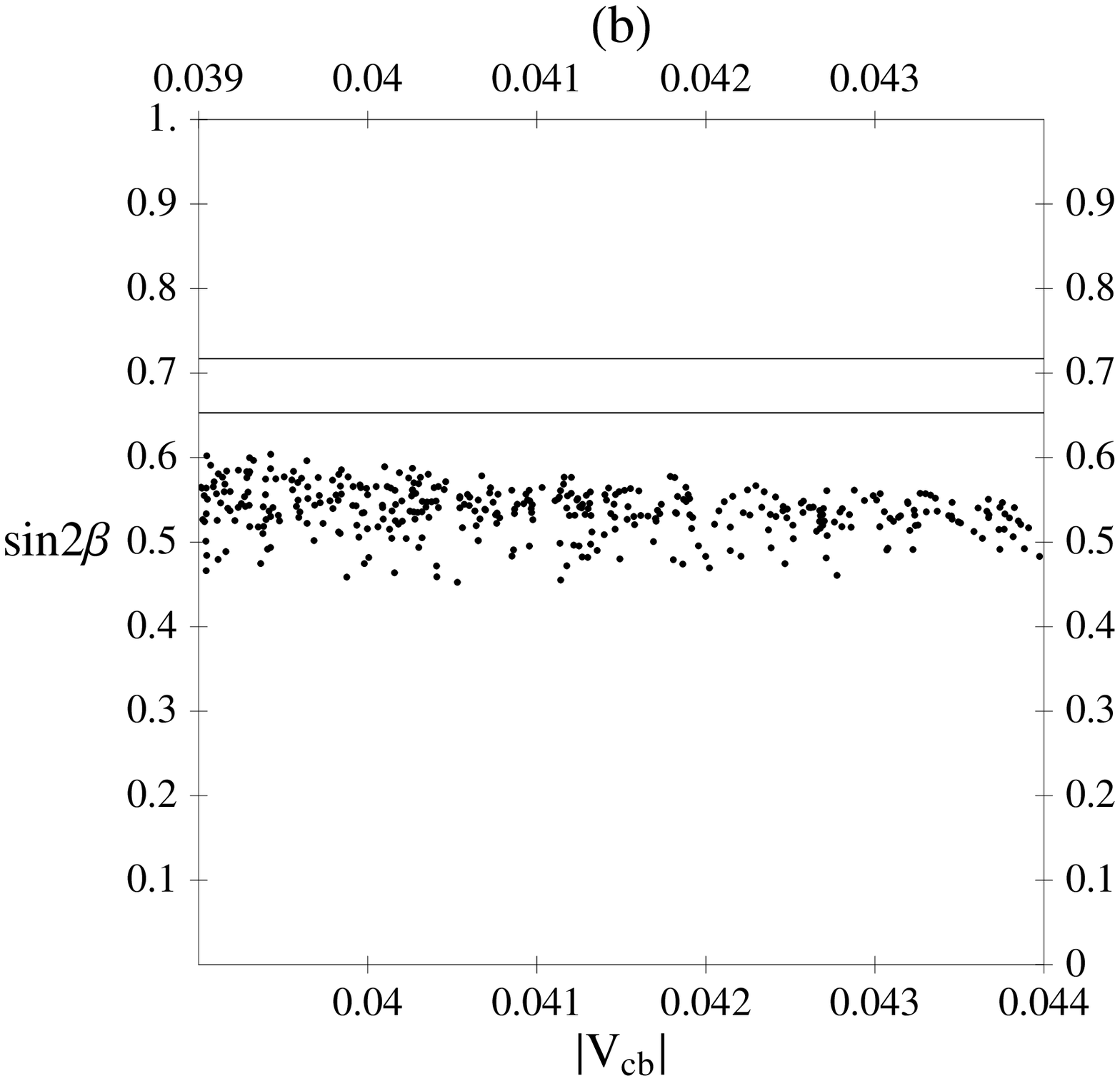}
}
\caption{(a): Predicted values of $\sin2\beta$ are plotted as a function of $|V_{ub}|$ for the case of U2-D2. 
(b): Predicted values of $\sin2\beta$ are plotted as a function of $|V_{cb}|$ for the case of U2-D2. 
In both figures, the constraints shown in eqs.(\ref{ex1}), (\ref{a3}) and (\ref{ex2}) are considered. 
The relevant parameters are fixed as $\tan\tilde{\beta}=50$, $x_{33}=1$, $M_{R1}=M_{R2}=10^{10}$ GeV and $M_{R3}=10^{15}$ GeV, respectively. 
The two horizontal lines in the figures correspond to the allowed range of $\sin2\beta$. 
}
\label{U2D2} 
\end{figure}

The recent measurements of the CP violating decay of B mesons into charmoniums provide us with the precise value of $\sin2\beta$~\cite{bccs}: 
\begin{eqnarray} 
\sin2\beta = 0.685\pm 0.032\ , 
\end{eqnarray} 
where $\beta$ is one of the angles in the CKM unitarity triangle. 
The angle $\beta$ is the most precisely measured quantity among the angles in the CKM unitarity 
triangle at present. 
The angle $\beta$ is given in terms of the elements in CKM matrix by   
\begin{eqnarray} 
\beta={\rm arg}\left(\frac{-V_{cd}V_{cb}^*}{V_{td}V_{tb}^*}\right)\ . 
\end{eqnarray} 
We also take the experimental allowed range of the CKM matrix elements \cite{PDG} : 
\begin{eqnarray} 
|V_{ub}|=0.0029-0.0045\ ,\ \ \ 
|V_{cb}|=0.039-0.044\ . 
\label{ex2} 
\end{eqnarray} 
It is obvious that the predictions of $\sin2\beta$ and the CKM matrix elements are closely related each other. 

Let us show the difference between the cases of U1 and U2 at first. 
Before showing the numerical results, 
we estimate the absolute values of the CKM matrix elements and $\sin2\beta$ by following the discussion 
presented in ref.~\cite{raby}. 
For the case of U1-D1, we obtain the following approximated relations : 
\begin{eqnarray} 
|V_{us}|\simeq \sqrt{\frac{m_d}{m_s}} 
\ ,\ \ \ 
\frac{|V_{ub}|}{|V_{cb}|}\simeq 
\sqrt{\frac{m_u}{m_c}} 
\ ,\ \ \ 
\beta\simeq \arg\left(1-\sqrt{\frac{m_u m_s}{m_c m_d}}e^{i\phi}\right) 
\ , 
\label{ckmu1d1} 
\end{eqnarray} 
where the CP violating phase $\phi$ is the combination of ones of the 
quark mass matrices. These approximate forms are given at the GUT scale. 
One expects almost constant ratio of $|V_{ub}|$ to $|V_{cb}|$ from 
eq.(\ref{ckmu1d1}) . 
For the case of U2-D1, we obtain the following relations : 
\begin{eqnarray} 
|V_{us}|\simeq \sqrt{\frac{m_d}{m_s}} 
\ ,\ \ \ 
|V_{ub}|\simeq 
\sqrt{\frac{m_u}{m_t}} 
\ ,\ \ \ 
\beta\simeq \arg\left(1-\sqrt{\frac{m_u m_s}{m_t m_d}}\frac{1}{|V_{cb}|}e^{i\phi'}\right) 
\ , 
\label{ckmu2d1} 
\end{eqnarray} 
where the CP violating phase $\phi'$ is also a combination of ones of 
the quark mass matrices. 
Comparing eq.~(\ref{ckmu1d1}) with eq.~(\ref{ckmu2d1}), 
one finds that the angle $\beta$ in the case of U2-D1 is predicted to be larger than the U1-D1 
and may be favored by experimental data. 

Fig. \ref{U1D1} shows the predicted values of $|V_{ub}|$, $|V_{cb}|$ and $\sin2\beta$ for the case 
of U1-D1 which are to be consistent with the constraints shown 
in eqs.~(\ref{ex1}) and (\ref{a3}). 
In all analyses of Fig. \ref{U1D1}, Fig.~\ref{U2D1} and Fig.~\ref{U2D2}, 
the relevant parameters are fixed as $\tan\tilde{\beta}=50$ and $x_{33}=1$, $M_{R1}=M_{R2}=10^{10}$ GeV 
and $M_{R3}=10^{15}$ GeV, respectively. 
The $|V_{ub}|$ is predicted to be near the lower bound in the case of U1-D1, and the predicted value 
of $|V_{cb}|$ is proportional to $|V_{ub}|$. 
One can also easily find that the predicted value of $\sin2\beta$ does not reach the experimental 
lower bound.   
Thus, we should exclude the case of U1 due to the observed value of $\sin2\beta$. 
Therefore, we will investigate only the case of U2 in more quantitative way in the following. 

Fig.~\ref{U2D1} shows the predicted values of the same parameter sets for the case of U2-D1 
in taking account of the constraints in eq.~(\ref{ex2}) in addition to eqs.~(\ref{ex1}) and (\ref{a3}). 
The $|V_{ub}|$ and $|V_{cb}|$ are found to cover the whole experimentally allowed region in this case. 
The predicted values of $\sin2\beta$ can reach the experimental allowed region only for $|V_{ub}|>0.0036$. 
We can conclude that there exists the allowed region where all 
the experimental constraints including $\sin2\beta$ are satisfied in the 
case of U2-D1. 

For the case of U2-D1$'$, we find the predicted values of the same parameter sets as in the case of U2-D1, which are shown in Fig. \ref{U2D2}. 
This case has the much smaller allowed region rather than the case of U2-D1. 
Since the $|V_{ub}|$ is constrained to have very small value, 
it seems to be difficult to reach the allowed region of $\sin2\beta$. 
The predictions for the all textures which we take are summarized in the Table 1. 
We can see that the case of U2-D1 is the most favored texture, 
especially if we 
compare the predicted values  of $|V_{ub}|$ 
and $\sin2\beta$ of the other cases. 

\begin{table}[t]%[htdp] 
\begin{center} 
\begin{tabular}{|c|c|c|c|c|c|c|c|c|c|c|} 
\hline 
& $m_d(M_Z)$ 
& $m_s(M_Z)$ 
& $m_b(M_Z)$ 
& $|V_{ub}|$ 
& $|V_{cb}|$ 
& $\sin2\beta$   
\\ \hline 
U1-D1 
& 
$3.7-5.0$ 
& 
$55-74$ 
& 
$3.0-3.2$ 
& 
$\sim 0.003$ 
& 
all 
& 
$<0.6$ 
\\ \hline 
U2-D1 
& 
$3.5-4.8$ 
& 
$55-77$ 
& 
$3.0-3.2$ 
& 
$0.0036<$ 
& 
all 
& 
$<0.8$ 
\\ \hline 
U2-D1$'$ 
& 
$3.2-3.4$ 
& 
$81-85$ 
& 
$3.0-3.2$ 
& 
$<0.0034$ 
& 
all 
& 
$\sim 0.5-0.6$ 

\\ \hline 
\end{tabular} 
\caption{The table shows the summarized predictions for the different textures in the case of $\tan\tilde{\beta}=50$, $x_{33}=1$, $M_{R1}=M_{R2}=10^{10}$ GeV and $M_{R3}=10^{15}$ GeV, respectively. 
The "all" means that the predicted region covers all the experimentally allowed region. } 
\end{center} 
\label{ts} 
\end{table} 

%%%%%%%%%%%%%%%%%%%%%%% 
%%%%%%%%%%%%%%%%%%%%%%% 
\section{Discussion} 
\label{dis} 
%%%%%%%%%%%%%%%%%%%%%%% 
%%%%%%%%%%%%%%%%%%%%%%% 

We have investigated the symmetric two-zero textures for all quark/lepton mass matrices in the SUSY SO(10) 
GUT, in which the Georgi-Jarlskog relations are realized. 
The low-energy predictions are obtained by solving the renormalization group equations with the right-handed neutrino threshold effects. 
One of the important consequence of such framework is that the bottom quark and tau lepton masses are 
unified at the GUT scale. 
The large $\tan\tilde{\beta}$ and the heaviest right-handed neutrino mass : $M_{R3}$ near the GUT scale are favored by the bottom-tau unification.   
The CG coefficient in 3-3 element of neutrino Dirac mass matrix : $x_{33}$ is also constrained to be $1$ not $-3$. 
The predicted value of $\sin2\beta$ strongly depends on the choice of up-quark mass matrix textures. 
As results of the analysis, 
the case of U2-D1 is favored by the measurement of $\sin2\beta$ while the other cases which include U1-D1, U1-D2 and U2-D2, are not. 
However, the allowed region on the $|V_{ub}|-\sin2\beta$ plane in the case of U2-D1 is very limited. 

Finally, 
we would like to discuss the possible textures and patterns of the CG coefficients for the neutrino Dirac 
mass matrix. 
These are directly related to neutrino oscillation observables. 
The threshold effect of the  lightest and the next-lightest 
right-handed neutrino masses : $M_{R1}$ and $M_{R2}$ in the RGEs running become 
much important to predict neutrino oscillation observables. 
We can obtain a strong constraint to the parameters in the neutrino Dirac mass matrix and 
the right-handed neutrino one from such analysis.%~\cite{BKOTf}. 
We can expect that such investigation leads us to a complete framework 
of matter unification.

\section*{Acknowledgements}
M.B thanks to M.C.Chen for useful discussion, especially for pointing out the 
importance of the observed quantities appearing in unitary triangle 
diagram. 
The work of S.K. has been supported 
by the Japan Society of Promotion of Science. 

%\appendix
%\section{First Appendix} %Empty argument \section{} yields `Appendix'. 
%
%\section{Second Appendix}


\begin{thebibliography}{99}

%%%%%%%%%%%%%%%%%%%%%%%%%%%%%%%%%%%%%%%%%%%%%%%%%%%%%%%%%%%%%
% Some macros are available for the bibliography:
%  o for general use
%    \JL : general journals                 \andvol : Vol (Year) Page
%  o for individual journal 
%    \AJ   : Astrophys. J.           \NC         : Nuovo Cim.
%    \ANN  : Ann. of Phys.           \NPA, \NPB  : Nucl. Phys. [A,B]
%    \CMP  : Commun. Math. Phys.     \PLA, \PLB  : Phys. Lett. [A,B]
%    \IJMP : Int. J. Mod. Phys.      \PRA - \PRE : Phys. Rev. [A-E]     
%    \JHEP : J. High Energy Phys.    \PRL        : Phys. Rev. Lett.
%    \JMP  : J. Math. Phys.          \PRP        : Phys. Rep.
%    \JP   : J. of Phys.             \PTP        : Prog. Theor. Phys.     
%    \JPSJ : J. Phys. Soc. Jpn.      \PTPS       : Prog. Theor. Phys. Suppl.
% Usage:
%  \PRD{45,1990,345}          ==> Phys.~Rev.\ \textbf{D45} (1990), 345
%  \JL{Nature,418,2002,123}   ==> Nature \textbf{418} (2002), 123
%  \andvol{B123,1995,1020}    ==> \textbf{B123} (1995), 1020
%%%%%%%%%%%%%%%%%%%%%%%%%%%%%%%%%%%%%%%%%%%%%%%%%%%%%%%%%%%%%
%  
%%%%%%%%%%%%%%%%%%%%%%% 
\bibitem{GJ} 
H. Georgi and C. Jarlskog, \PLB{86,1979,297}. 
%%%%%%%%%%%%%%%%%%%%%%% 

%%%%%%%%%%%%%%%%%%%%%%% 
\bibitem{RRR} 
P. Ramond, R.G. Roberts and G.G. Ross, \NPB{406,1993,19}. 
%%%%%%%%%%%%%%%%%%%%%%% 

%%%%%%%%%%%%%%%%%%%%%%% 
\bibitem{watanabe} 
N. Uekusa, A. Watanabe and K. Yoshioka, \PRD{71,2005,094024}. 
%%%%%%%%%%%%%%%%%%%%%%% 

%%%%%%%%%%%%%%%%%%%%%%% 
\bibitem{raby} 
H.D. Kim, S. Raby and L. Schradin, \PRD{69,2004,092002}. 
%%%%%%%%%%%%%%%%%%%%%%% 

%%%%%%%%%%%%%%%%%%%%%%% 
\bibitem{btau} 
F. Vissani and A.Yu. Smirnov, \PLB{341,1994,173}; 
A. Brignole, H. Murayama and R.\ Rattazzi, \PLB{335,1994,345}; 
M.  Bando and K. Yoshioka, \PLB{444,1998,373}; 
A. Kageyama, M. Tanimoto and K. Yoshioka, \PLB{512,2001,349}. 
%%%%%%%%%%%%%%%%%%%%%%% 

%%%%%%%%%%%%%%%%%%%%%%% 
\bibitem{BO} 
M. Bando and M. Obara, \PTP{109,2003,995}. 
%%%%%%%%%%%%%%%%%%%%%%% 

%%%%%%%%%%%%%%%%%%%%%%% 
\bibitem{BKOT1} 
M. Bando, S. Kaneko, M. Obara and M. Tanimoto, \PLB{580,2004,229}. 
%%%%%%%%%%%%%%%%%%%%%%% 

%%%%%%%%%%%%%%%%%%%%%%% 
\bibitem{BKOT2} 
M. Bando, S. Kaneko, M. Obara and M. Tanimoto, \PTP{112,2004,533}. 
%%%%%%%%%%%%%%%%%%%%%%% 

%%%%%%%%%%%%%%%%%%%%%%% 
\bibitem{Chen} 
M.C. Chen and K.T. Mahanthappa, \PRD{62,2000,113007}; 
M.C. Chen and K.T. Mahanthappa, \PRD{65,2002,053010}; 
M.C. Chen and K.T. Mahanthappa, \PRD{68,2003,017301}; 
M.C. Chen and K.T. Mahanthappa, \PRD{70,2004,113013}. 
%%%%%%%%%%%%%%%%%%%%%%% 

%%%%%%%%%%%%%%%%%%%%%%% 
\bibitem{two-zero3} 
H. Nishiura, K. Matsuda and T. Fukuyama, \PRD{60,1999,013006}; 
K. Matsuda, T. Fukuyama and H. Nishiura, \PRD{61,2000,053001}. 
%%%%%%%%%%%%%%%%%%%%%%% 

%%%%%%%%%%%%%%%%%%%%%%%% 
\bibitem{FuYa} 
%%%%%%%%%%%%%%%%%%%%%%%% 
M. Fukugita and T. Yanagida, 
\PLB{175,1986,45}. 

%%%%%%%%%%%%%%%%%%%%%%%% 
\bibitem{pilaftsis} 
%%%%%%%%%%%%%%%%%%%%%%%% 
A.~Pilaftsis, 
\PRD{56,1997,5431}; 
%``Heavy Majorana neutrinos and baryogenesis,'' 
\IJMP{A14,1999,1811};
A. Pilaftsis and Thomas E.J. Underwood, 
\NPB{692,2004,303}. 

%%%%%%%%%%%%%%%%%%%%%%% 
\bibitem{RGEs} 
S. Antusch, J. Kersten, M. Lindner and M. Ratz, 
\PLB{538,2002,87}; 
S. Antusch, J. Kersten, M. Lindner and M. Ratz, 
\NPB{674,2003,401}; 
S. Antusch, J. Kersten, M. Lindner, M. Ratz and M.A. Schmidt, 
\JHEP{0503,2005,024}.   
%%%%%%%%%%%%%%%%%%%%%%% 

%%%%%%%%%%%%%%%%%%%%%%% 
\bibitem{PDG} 
S. Eidelman et al., \PLB{592,2004,1}. 
%%%%%%%%%%%%%%%%%%%%%%% 

%%%%%%%%%%%%%%%%%%%%%%% 
\bibitem{KF} 
H. Fusaoka and Y. Koide, 
\PRD{57,1998,3986}.   
%%%%%%%%%%%%%%%%%%%%%%% 

%%%%%%%%%%%%%%%%%%%%%%% 
\bibitem{bccs} 
T. Hara, Talk at Flavor Physics and CP Violation Conference (FPCP 2006). 
%%%%%%%%%%%%%%%%%%%%%%% 


\end{thebibliography}
\end{document}